# Machine learning guided discovery of superconducting calcium borocarbides

Chao Zhang,[1] Hui Tang, [1] Chen Pan, [1] Hong Jiang [1] Huai-Jun Sun,[2] Kai-Ming Ho,[3] and Cai-Zhuang Wang[3.4*]

[1] *Department of Physics, Yantai University, Yantai 264005, China*

[2] *Jiyang College of Zhejiang Agriculture and Forestry University, Zhuji, 311800, China*

[3] *Department of Physics and Astronomy, Iowa State University, Ames, Iowa 50011, United States*

[4] *Ames National Laboratory, Ames, Iowa 50011, United States*

Pursuit of superconductivity in light-element systems at ambient pressure is of great experimental and theoretical interest. In this work, we combine a machine learning (ML) method with first-principles calculations to efficiently search for the energetically favorable ternary Ca-B-C compounds. Three new layered borocarbides (stable $CaBC_5$ and metastable $Ca_2BC_{11}$ and $CaB_3C_3$) are predicted to be phonon-mediated superconductors at ambient pressure. The hexagonal $CaB_3C_3$ possesses the highest $T_c$ of 26.05 K among the three compounds. The σ-bonging bands around the Fermi level account for the large electron-phonon coupling ($\lambda$ = 0.980) of hexagonal $CaB_3C_3$. The ML-guided approach opens up a way for greatly accelerating the discovery of new high-$T_c$ superconductors.

# 1. Introduction

Migdal-Eliashberg phonon mediated theory for superconductivity suggests that light-element compounds would be promising candidates for superconductors with high transition temperature ($T_c$). In contrast to heavy-element compounds in which the energy scale of phonons is usually much smaller than that of electrons, light-element materials tend to have high-frequency phonons (owing to their light atomic mass) and offer a unique playground for nontrivial interplay between the Coulomb correlations and electron-phonon interactions.[1, 2] Pressurized hydrides have been predicted and observed to have high $T_c$ above 200 K, however, the high-$T_c$ hydrides are synthesized and stabilized at high pressures (> 100 GPa).[3] For example, high-$T_c$ superconductivity was established above 170 and 166 GPa for clathrate metal hydrides La-H and Y-H system,[4-6] and near 155 GPa for $H_3S$.[7] Since the requirement of very high pressures would hinder the practical applications, the pursuit of high-$T_c$ superconductor that can persist in stable or metastable compounds at lower, and even ambient, pressure has generated a considerable recent intertest, yet remains an outstanding challenge.

Boron and/or carbon compounds would be good candidates for high-$T_c$ superconductor at ambient pressure. The well-known superconducting $MgB_2$ possesses a $T_c$ of 39 K,[8] due to strong coupling between σ-bonding electrons and B-B in-plane stretching vibrational phonons.[9-14] Superconductivity was also found in graphite intercalated compounds (GICs), but the $T_c$ are generally low (< 2 K),[15] expect $CaC_6$ which exhibits the highest $T_c$ of 11.5 K in this class of materials.[16, 17] Isovalent with and structurally similar to $MgB_2$, hole-doped layered $Li_xBC$ was suggested as a high-$T_c$ superconductor.[18, 19] However, superconducting $Li_xBC$ has not been observed in experiments due to doping caused strong lattice distortion and considerable changes in electronic band structures.[20-23] It was thus proposed that substituting carbon atoms with boron atoms would introduce hole doping in LiBC while retain the stability of the lattice. First-principles calculations suggested that layered $Li_3B_4C_2$, $Li_2B_3C$, $LiB_{1.1}C_{0.9}$, and $Li_4B_5C_3$ are superconductors with strong electron-phonon coupling. Moreover, alkaline earth metal intercalated layered compounds XBC (X = Mg, Ca, Sr, Ba) which adopt LiBC structure[24] and layered $CaB_3C_3$ which is structurally similar to $CaC_6$,[25] were also predicted to be phonon-mediated superconductors.

Apart from layered metal-intercalated borocarbide compounds, superconductivity was also found in carbon-boron clathrates. A carbon-boron clathrate $SrB_3C_3$, which was successfully synthesized at a pressure of 57 GPa[26], was theoretically predicted to be a superconductor with $T_c \sim 40$ K.[27, 28] This clathrate is composed of a single truncated octahedral B-C cage with Sr atoms incorporated at the void of the B-C cage. The electron-phonon coupling in clathrate $SrB_3C_3$ comes from the $sp^3$-hybridized σ-bands and the boron-associated $E_g$ phonon modes. The $T_c$ was further enhanced to 75 K when partially substituting Sr with Rb.[29, 30] By replacing Sr atom in the clathrate $SrB_3C_3$ with the first 57 elements (Z = 1 – 57), $CaB_3C_3$, $YB_3C_3$, and $LaB_3C_3$ clathrates were predicted to be stable at ambient pressure by a high-throughput first-principles calculations.[30]

There is only one stoichiometric calcium borocarbide, $CaB_2C_2$, synthesized so far for the Ca-B-C ternary system. The crystal structure of $CaB_2C_2$ was firstly identified to be isostructural to $LaB_2C_2$.[31] Subsequent studies using X-ray diffraction method proposed two structures with space groups, *I4/mcm* and *P4/mbm*, respectively, would be the ground state of $CaB_2C_2$.[32, 33] In fact, the total energies of the two structures are very similar using first-principles calculations.[33] The *I*4/*mcm* phase was finally determined to be the ground state of $CaB_2C_2$ using electron energy-loss spectroscopy.[34] However, neither *I*4/*mcm* nor *P*4/*mbm* phase of $CaB_2C_2$ is a superconductor. Whether there are stable compounds with high-$T_c$ in the Ca-B-C system at ambient pressure remains an open question.

In this work, we take the Ca-B-C ternary system as a prototype alkaline earth borocardies to predict new ternary compounds using an efficient framework which integrates machine learning (ML) and first-principles calculations.[35, 36] Three stable and three low-energy metastable structures of calcium borocarbide are found and three of them ($CaBC_5$, $Ca_2BC_{11}$, and $CaB_3C_3$) are predicted to be phonon-mediated superconductors.

## 2. Computational details

The low-energy structures and compositions of Ca-B-C system are explored using an interactive framework which combines an efficient ML model for high-throughput screening and first-principles methods for accurate calculations. A crystal graph conventional neural network (CGCNN) ML model [37] is employed to perform the fast high-throughput screening over a wide range of possible compositions and crystal structures to select promising candidates for low-energy Ca-B-C ternary compounds.

We construct a hypothetical structure pool for ternary Ca-B-C compounds by collecting 11,914 known ternary structures from the MP database[38] and replace the three elements with Ca, B, and C. For a given ternary structure from the MP database, five hypothetical lattices are generated by uniformly changing the original volume by a factor of 0.96—1.04 with an interval of 0.02. There are six ways to shuffle the three elements on a given template ternary structure. Thus, a set of 357,420 hypothetical Ca-B-C structures are generated based on MP database (we refer to these structures as MP-based structures). We also generate another set of 65,287 structures using a random generation algorithm (refer to as RS-based structures).[39]

The CGCNN model for predicting the formation energies of compounds developed by Xie and Grossman in Ref. [37] was trained using the DFT calculated structures and energies of 28,046 binary and ternary compounds involving a wide range of chemical elements in the Materials Project (MP) database. We refer to this model as the first generation (1G-CGCNN) model. By applying the 1G-CGCNN model to the MP-based and RS-based Ca-B-C ternary structures generated above, we select only 1200 and 1200 structures with negative formation energies from MP-based and RS-based methods, respectively, for subsequent first-principles calculations. The results of the first-principles calculation on these 2400 candidate structures will be used to retrained the CGCNN model (refer to as 2G-CGCNN model) to improve the accuracy specifically for Ca-B-C system.

In order to generate sufficient data for training the 2G-CGCNN model, we also performed high-throughput first-principles calculations by substituting alkali and alkaline metals by Ca in the known stable alkali and alkaline metal borides, carbides, and borocarbides. We found some stable and metastable Ca-B binary, Ca-C binary and Ca-B-C ternary compounds, i.e, $P6_3/mmc$-$CaB_2$, $P4/mbm$-$CaB_3$, $Imma$-$CaB_7$, $R\bar{3}m$-$CaC_2$, $P6/mmm$-$CaC_8$, $P6/mmm$-$CaC_{12}$, $Imma$-$CaB_{12}C_2$, and $P\bar{4}n2$-$Ca_2B_{24}C$. For each crystalline structure of the stable and metastable of Ca-B, Ca-C, and Ca-B-C compounds, we uniformly expand or contract the unit cell of the structure by a scaling factor ranging from 0.85 to 1.25 with 0.05 interval, and at the same time the atomic positions in each unit cell are perturbed randomly for 50 times with distortion amplitudes in the range from −0.025 to 0.025 times the length of the cell vector.

In this way, 4896 distorted crystalline structures are generated (refer to as DC-based structures) and the total energies of the structures are calculated by first-principles calculations.

Based on the first-principles calculation results on the 7296 structures from the three different generation methods discussed above, we select 6271 structures whose formation energies are smaller than 2 eV/atom to train the 2G-CGCNN model specifically for Ca-B-C system. We then apply the 2G-CGCNN ML model to the set of 357,420 MP-based structures and to the set of 65,287 RS-based structures. A total of 2000 structures with low formation energy predicted by the 2G-CGCNN model from the structure pool generated by MP and RS methods are then selected for the structure relaxation by first-principles calculations. The distribution of the formation energies of hypothetical structures predicted by the 1G- and 2G-CGCNN models are given in the Supplemental Material.

The first-principles calculations are carried out according to density functional theory (DFT) within the framework of the all-electron projector-augmented wave (PAW) method,[40] as implemented in the Vienna Ab inito Simulation Package (VASP).[41] We adopt the Perdew-Burk-Ernzerhof (PBE) functionals at the generalized gradient approximation (GGA) level.[42] A plane wave basis set with a kinetic energy cut-off of 520 eV is used and a uniform, $\Gamma$-centered $k$ mesh with $2\pi \times 0.03$ Å$^{-1}$ spacing. The electron–phonon coupling (EPC) is calculated using the Quantum-Espresso (QE) code[43] with the PBE functional and PAW pseudopotentials. The planewave basis set and the charge density are expanded with the kinetic energy cut-offs of 60 and 600 Ry, respectively, in the EPC calculations.

## 3. Results and discussion

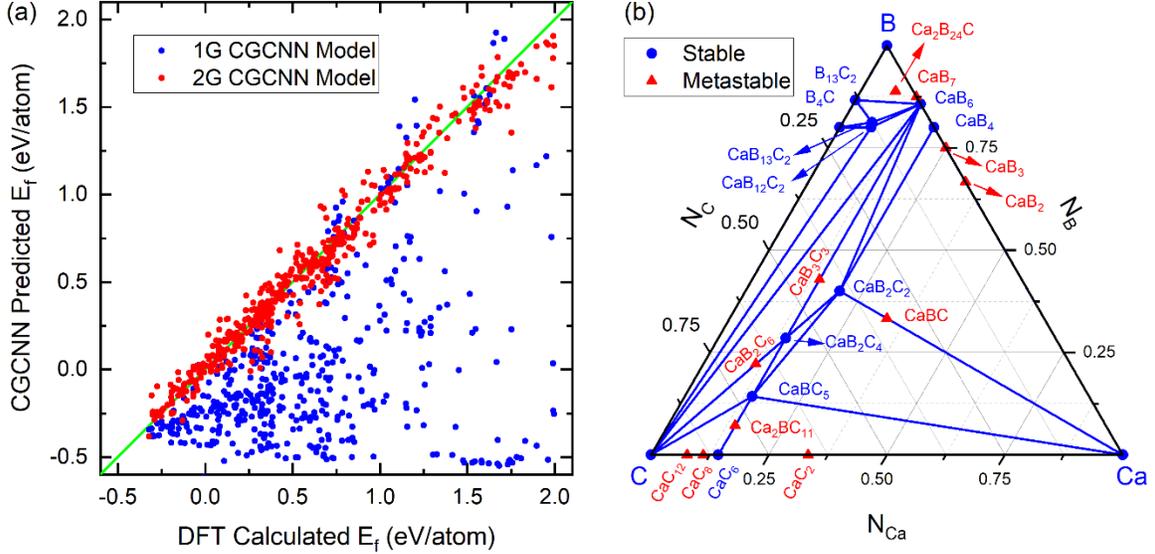

Figure 1. (a) Formation energy of Ca-B-C system predicted by CGCNN models are compared with those from DFT. (b) Ternary phase diagram of Ca-B-C system at ambient pressure.

In order to assess the reliability of the CGCNN predictions, we random select 500 structures from the train database, and compare the predicted formation energies from the 1G- and 2G-CGCNN ML models with those from first-principles calculations. The performance of the 2G-CGCNN ML model is significantly better than that of the 1G-CGCNN model, as shown in Fig. 1(a). The 1G-CGCNN ML model significantly underestimates the formation energies of the Ca-B-C system. The root mean square error (RMSE) of the 2G-CGCNN model is 0.092 eV/atom, which is much smaller than 0.765 eV/atom from the 1G-CGCNN model. Thus, our trained 2G-CGCNN model is more suitable for the Ca-B-C system.

The 2G-CGCNN model enable us to discover two stable calcium borocarides, $CaBC_5$ with *Amm*2 symmetry and $CaB_{13}C_2$ with *C*2/*m* symmetry. Based on $CaBC_5$, we build $Ca_2BC_{11}$ with *Amm*2 symmetry, $CaB_2C_4$ with *Amm*2 symmetry, and $CaB_3C_3$ with $P\bar{6}2m$ symmetry. The structural relationships between CaBC$_5$, Ca$_2$BC$_{11}$, CaB$_2$C$_4$, and CaB$_3$C$_3$ will be discussed below. The ternary phase diagram of the Ca-B-C system at ambient pressure is constructed using previously reported experimental structures and our predicted structures, as shown Fig. 1(b). *Amm*2-CaBC$_5$, *Amm*2-CaB$_2$C$_4$, *I*4/*mcm*-CaB$_2$C$_2$, *Imma*-CaB$_{12}$C$_2$, and *C*2/*m*-CaB$_{13}$C$_2$ are stable ternary compounds. *Amm*2-Ca$_2$BC$_{11}$, *Fmm*2-CaB$_2$C$_6$, $P\bar{6}2m$-CaB$_3$C$_3$, $P\bar{4}n2$-Ca$_2$B$_{24}$C, *P*6$_3$/*mmc*-CaBC are metastable ternary compounds, whose formation energies are 9.5 meV/atom, 41.6 meV/atom, 153.0 meV/atom, 168.6 meV/atom, and 201.4 meV/atom, respectively, above the convex hull. To gain insights into the effects of temperature on the stable and metastable compounds, we evaluated the Gibbs free energy as a function of temperature, as shown in Supplemental Material. The ternary phase diagram of the Ca-B-C system changes slightly below 500, and we selected a phase diagram at 300 K shown in Supplemental Material.

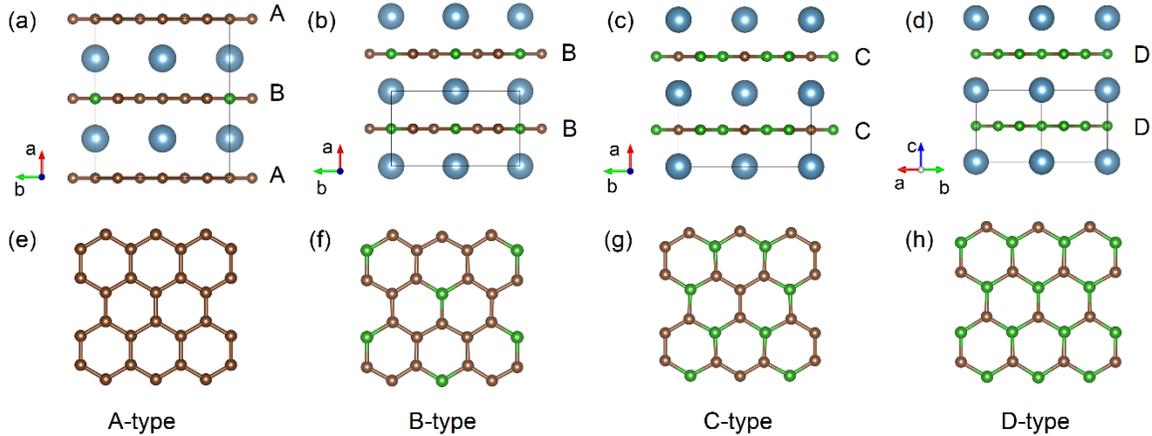

Figure 2. Side view of crystal structures of (a) Ca$_2$BC$_{11}$, (b) CaBC$_5$, (c) CaB$_2$C$_4$, and (d) CaB$_3$C$_3$. (e) Graphene layer and (f)-(h) boron-carbon layers. Dark blue, green, and brown balls represent Ca, B, and C atoms, respectively.

Figure 2 shows the crystal structure of $Ca_2BC_{11}$, $CaBC_5$, $CaB_2C_4$, and $CaC_3C_3$. Similar to GICs, these four compounds are Ca intercalated layered structures. Carbon and boron atoms form three types of graphene-like layers, which are termed B-type (denoted as B), C-type (denoted as C), and D-type (denoted as D) carbon-boron layer with increasing boron content, as shown in Figs. 2(f)-(h). The pristine graphene layer is termed A-type layer (denoted as A), as shown in Fig. 1(e). The intercalant Ca atoms form a triangular array between graphene layers or graphene-like carbon-boron layers, and each layer of Ca intercalation is denoted as α. The stacking sequences of the $Ca_2BC_{11}$, $CaBC_5$, $CaB_2C_4$, and $CaB_3C_3$ are thus in the pattern of AαBα, BαBα, CαCα, and DαDα, respectively. In this way, $Ca_2BC_{11}$, $CaBC_5$, and $CaB_2C_4$ adopt orthorhombic structures with *Amm*2 symmetry, whereas $CaB_3C_3$ form a hexagonal structure with $P\bar{6}2m$ symmetry.

Instead of being exactly in the middle of the adjacent graphene layer and boron-carbon layer, Ca atoms are biased towards the graphene layer. With increasing content of boron in $Ca_2BC_{11}$, $CaBC_5$, and $CaB_2C_4$, the interlayer distance between Ca layer and boron-carbon layer decreases. As a result, the average C-C and B-C bond lengths in boron-carbon layer slightly increase from $Ca_2BC_{11}$, $CaBC_5$, to $CaBC_4$. For the highest ratio of boron to carbon (1:1) in $CaB_3C_3$, the average B-C bond length is 1.566 Å. The average C-C bond length in the graphene layer of $Ca_2BC_{11}$ is 1.467 Å, which is slightly larger than that of 1.449 Å in $CaC_6$. It is noteworthy that the stoichiometric $CaB_3C_3$ with $P\bar{6}2m$ symmetry predicted in this work has very similar energy with another $CaB_3C_3$ with *R*32 symmetry which was proposed by Chen.[25] However, the symmetry of $P\bar{6}2m$-$CaB_3C_3$ is higher than that of *R*32-$CaB_3C_3$. In addition, the metastable $Ca_2B_{24}C$, stable $CaB_{13}C_2$, and stable $CaB_{12}C_2$ contain B cages, and details of structural properties of these compounds are shown in the Supplemental Material.

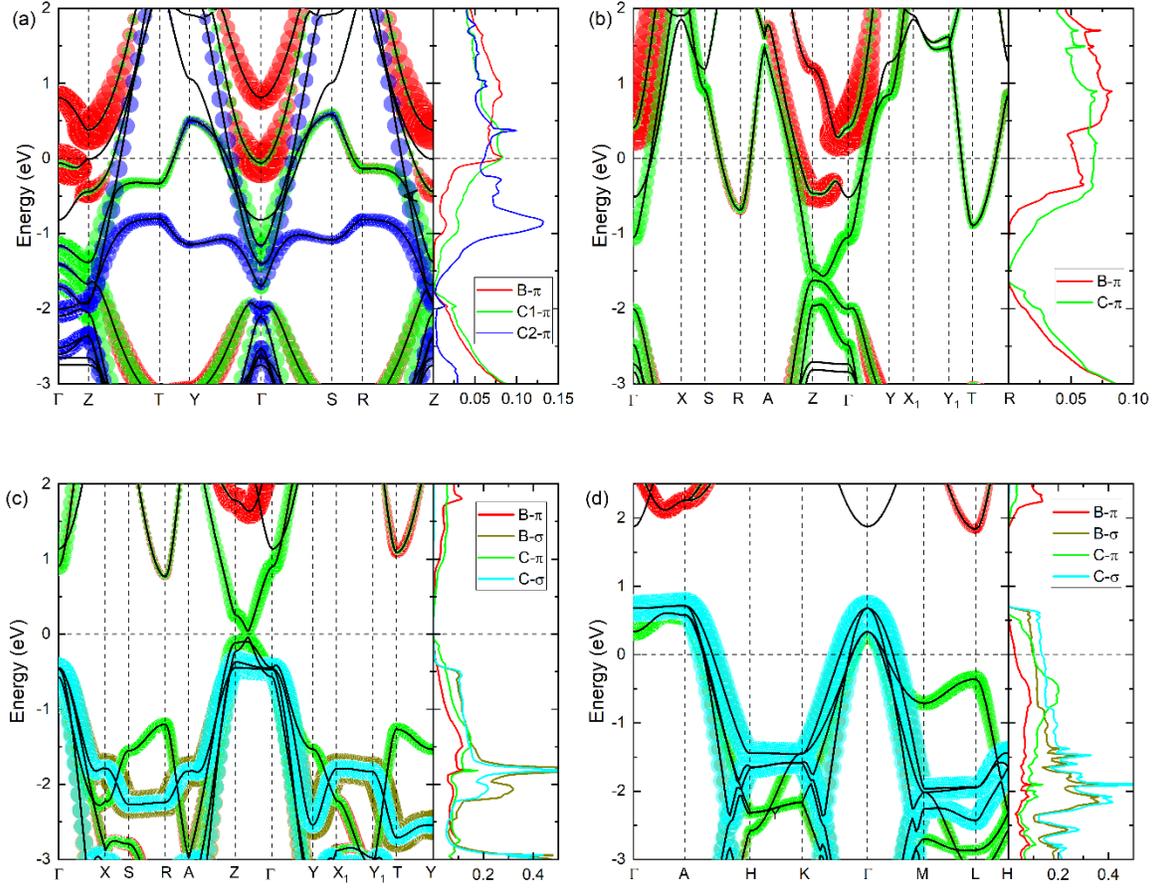

Figure 3. Orbital-resolved electronic band structure and projected density of states (PDOS) of (a) $Ca_2BC_{11}$, (b) $CaBC_5$, (c) $CaB_2C_4$, and (d) $CaB_3C_3$. The unit of electronic PDOS is states/eV/atom.

The orbital-resolved electronic band structures and projected density-of-states (PDOSs) of $Ca_2BC_{11}$, $CaBC_5$, $CaB_2C_4$, and $CaB_3C_3$ are plotted in Fig. 3. The $Amm2$-$Ca_2BC_{11}$, $Amm2$-$CaBC_5$, and $P\bar{6}2m$-$CaB_3C_3$ are metallic, whereas the $Amm2$-$CaB_2C_4$ is a semiconductor. We focus on the electronic contributions of boron and carbon atoms around the Fermi level, and the total DOS and PDOS of Ca are shown in the Supplemental Material. For the $Amm2$- $Ca_2BC_{11}$ and $Amm2$-$CaBC_5$, the π electrons from B and C atoms play a vital role at the Fermi level compared with σ electrons from them. As one can see from Fig. 3(a) and 3(b), the π electrons from both boron and carbon in these two structures provide significant DOS at the Fermi level. The PDOS from the π electrons in mixed B-C layer decreases quickly below the Fermi level and reach almost zero at about 1.5 eV below the Fermi level, while the PDOS of the π electrons from the graphene layer in the $Amm2$-$Ca_2BC_{11}$ increases and forms a peak around 1 eV below the Fermi level combined with Ca $d$ states, as shown in Supplemental Material. Different from $Amm2$-$Ca_2BC_{11}$ and $Amm2$-$CaBC_5$, the σ electrons from B and C atoms in the $Amm2$-$CaB_2C_4$ and $P\bar{6}2m$-$CaB_3C_3$ structures play an important role in the PDOS around the Fermi level. The top of the valence band and the bottom of the conduction band of $Amm2$-$CaB_2C_4$ locate between the Z → Γ line, which results to a small gap of 0.083 eV at the PBE-level. The $P\bar{6}2m$-$CaB_3C_3$ is metallic with strong contributions from the σ electrons of boron and carbon atoms to the states at the Fermi level in addition to the π electrons contributions, although there is a band gap of ~1 eV starts from about 0.7 eV above the Fermi level. The band structure of this structure exhibits a combination of some flat bands and some steep bands in the vicinity of the Fermi level along the Γ → A direction. The simultaneous occurrence of flat and steep bands near the Fermi level has been suggested as essential to superconducting behavior. In addition, the B-rich ternary compounds ($Imma$-$CaB_{12}C_2$, $C2/m$-$CaB_{13}C_2$, and $P\bar{4}n2$-$Ca_2B_{24}C$) show semiconducting characteristics. The band gaps of $C2/m$-$CaB_{13}C_2$, $Imma$-$CaB_{12}C_2$, and $P\bar{4}n2$-$Ca_2B_{24}C$ are 0.799, 1.406, and 1.822 eV, respectively. Details of the electronic band structures of newly predicted B-rich compounds are shown in Supplemental Material.

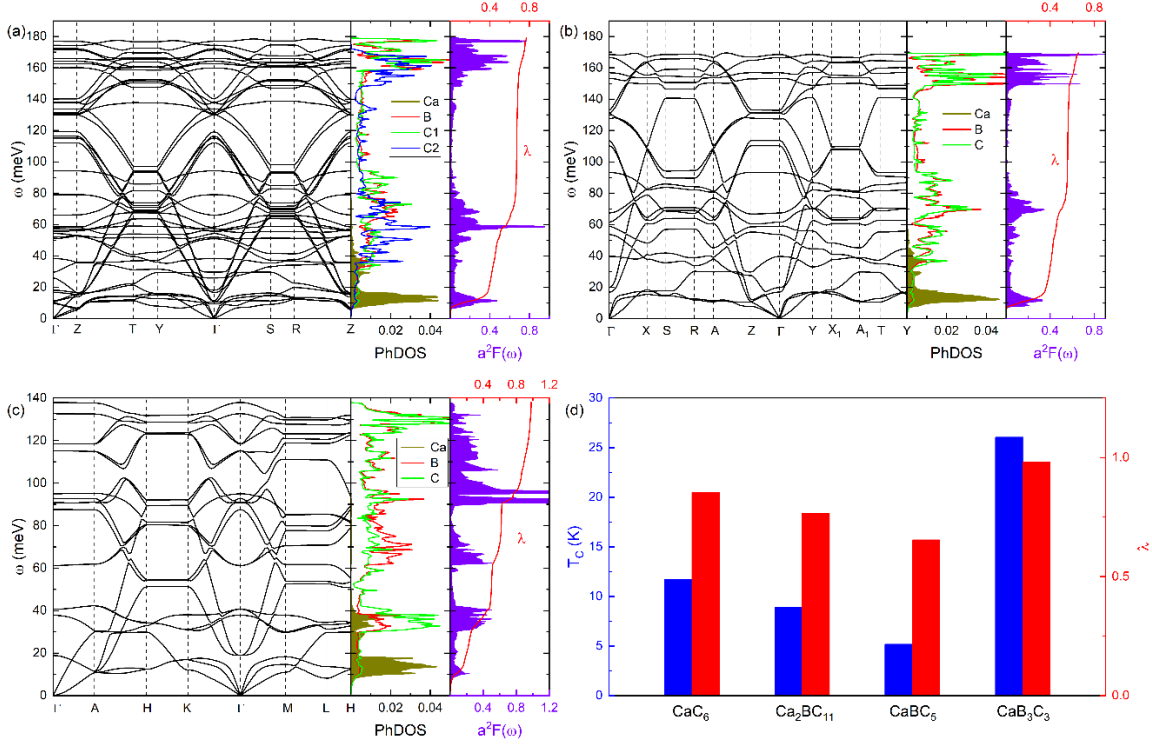

Figure 4. Phonon band structure, phonon PDOS, $\alpha^2F(\omega)$, and frequency-dependent integrated $\lambda(\omega)$ of (a) $Ca_2BC_{11}$, (b) $CaBC_5$, and (c) $CaB_3C_3$. (d) $T_c$ and $\lambda$ of $CaC_6$, $Ca_2BC_{11}$, $CaBC_5$, and $CaB_3C_3$.

Figure 4 presents the phonon band structure, phonon PDOS, Eliashberg spectral function $\alpha^2F(\omega)$, and integrated electron-phonon coupling constant $\lambda(\omega)$ of the $Amm2$-$Ca_2BC_{11}$, $Amm2$-$CaBC_5$, and $P\bar{6}2m$-$CaB_3C_3$. The absence of imaginary of phonon frequencies establishes the dynamical stabilities of these predicted structures. For the three compounds, the low-frequency range (below 20 meV) is dominated by Ca atoms and the vibration of Ca atoms extend to 60 meV. The frequency range above 60 meV is dominated by B and C atoms. The $Amm2$-$Ca_2BC_{11}$ and $Amm2$-$CaBC_5$ have similar distributions of Elaishberg spectral function, which lead to three steps in the integrated $\lambda(\omega)$. Vibration of Ca atoms mainly contribute to the first step of the integrated $\lambda(\omega)$, which are 51.8% and 55.6% of the total $\lambda$ for $Amm2$-$Ca_2BC_{11}$ and $Amm2$-

CaBC$_5$, respectively. The intermediate-frequency range (20-100 meV) and high-frequency range (150-180 meV) approximately contribute to 30% and 10% the total λ, respectively. The distribution of Elaishberg spectral function of the $P\bar{6}2m$-CaB$_3$C$_3$ is apparently different from that of the $Amm2$-Ca$_2$BC$_{11}$ and $Amm2$-CaBC$_5$. There are four steps in the integrated λ(ω). The low-frequency range (0-20 meV) dominated by Ca atoms only contributes 20.2% to the total λ. The frequency range located at 20-60 meV, which reflects strong interaction between Ca, B, and C atoms, contributes 31.9% to the total λ. The high-frequency range (80-140 meV) coming from B and C atoms contribute 36.9% to the total λ. Especially, the phonon vibration around 90 meV strongly couples with σ electrons of B and C atoms.

According to the Allen-Dynes modified version of the McMillan equation:[44, 45]

$$T_c = \frac{\omega_{log}}{1.2} \exp\left[-\frac{1.04(1+\lambda)}{\lambda-\mu^*(1+0.62\lambda)}\right],$$

we estimate the $T_c$ of Ca-B-C system at ambient pressure with the Coulomb pseudopotential $\mu^*$ of 0.15, as shown in Fig. 4(d). The predicted $T_c$ of $R\bar{3}m$-CaC$_6$ is 11.7 K which is in excellent consistent with the experimental data (11.5 K).[16] With increasing of B contents from the $R\bar{3}m$-CaC$_6$, the $Amm2$-CaBC$_{11}$, to the $Amm2$-CaBC$_5$, the λ monotonically decreases from 0.853 to 0.766 and to 0.655, Correspondingly, $T_c$ decreases from 11.75 K to 8.92 K and to 5.19 K. This comes from the similar Eliashberg distributions of these three compounds. $P\bar{6}2m$-CaB$_3$C$_3$ possesses the largest $T_c$ (26.05 K) among the four compounds. The high $T_c$ stems from the strong coupling between σ-bonding and B and C vibration in the frequency range of 87 and 95 meV. It is worth noting that the $T_c$ of $R32$-CaB$_3$C$_3$ is estimated to be 26.73 K, which is agreement with $T_c$ of 28.2 K obtained by Chen.[25]

## 4. Conclusion

In summary, we search for low-energy Ca-B-C ternary compounds using an efficient framework which combines ML high-throughput screening with accurate first-principles calculations, and explore the possible superconductivity of these new compounds at ambient pressure. Four new stable ($Amm2$-CaBC$_5$, $Amm2$-CaB$_2$C$_4$, $Imma$-CaB$_{12}$C$_2$, and $C2/m$-CaB$_{13}$C$_2$) and three new metastable ($Amm2$-Ca$_2$BC$_{11}$, $P\bar{6}2m$-CaB$_3$C$_3$, and $P\bar{4}n2$-Ca$_2$B$_{24}$C) calcium borocarbides are revealed. Layered $Amm2$-Ca$_2$BC$_{11}$, $Amm2$-CaBC$_5$ and $P\bar{6}2m$-CaB$_3$C$_3$ are predicted to be phonon-mediated superconductors, and $P\bar{6}2m$-CaB$_3$C$_3$ possesses the highest $T_c$ of 26.05 K among the three compounds. $Amm2$-CaB$_2$C$_4$, $C2/m$-CaB$_{13}$C$_2$, $Imma$-CaB$_{12}$C$_2$, and $P\bar{4}n2$-Ca$_2$B$_{24}$C are semiconductors, which have band gaps of 0.083 eV, 0.799 eV, 1.406eV, and 1.822 eV, respectively. The stable and metastable structures of the Ca-B-C system have significant implications for alkali and alkaline metal borocarbides. The ML-guided approach opens up a way for greatly accelerating the discovery of new high-$T_c$ superconductors.

## Acknowledgements

C. Zhang was supported by the National Natural Science Foundation of China (Grants No. 11874318). Work at Ames National Laboratory was supported by the U.S. Department of Energy (DOE), Office of Science, Basic Energy Sciences, Materials Science and Engineering Division including a grant of computer time at the National Energy Research Supercomputing Center (NERSC) in Berkeley. Ames National Laboratory is operated for the US DOE by Iowa State University under Contract No. DEAC02-07CH11358.